\documentclass[twocolumn,showpacs,preprintnumbers,amsmath,amssymb]{revtex4}
\usepackage{graphicx}% Include figure files
\usepackage{dcolumn}% Align table columns on decimal point
\usepackage{bm}% bold math

\begin{document}

\preprint{}

\title{Radiative charge transfer lifetime\\ of the excited state of
  (NaCa)$^+$}% Force line breaks with \\

\author{Oleg P. Makarov}%
\altaffiliation{}
 \email{Oleg.Makarov@UConn.edu}
\author{R. C\^ot\'e}%
\author{H. Michels}
\author{W. W. Smith}
\affiliation{%
Department of Physics, University of Connecticut, Storrs, CT, 06269-3046
}%

\date{\today}% It is always \today, today,

\begin{abstract}
  New experiments were proposed recently to investigate the regime of
  cold atomic and molecular ion-atom collision processes in a special
  hybrid neutral-atom--ion trap under high vacuum conditions. The
  collisional cooling of laser pre-cooled Ca$^+$ ions by ultracold Na
  atoms is being studied. Modeling this process requires knowledge of
  the radiative lifetime of the excited singlet A$^1\Sigma^+$ state of
  the (NaCa)$^+$ molecular system. We calculate the rate coefficient
  for radiative charge transfer using a semiclassical approach. The
  dipole radial matrix elements between the ground and the excited
  states, and the potential curves were calculated using Complete
  Active Space Self-Consistent field and M\"oller-Plesset second order
  perturbation theory (CASSCF/MP2) with an extended Gaussian basis,
  6-311+G(3df). The semiclassical charge transfer rate coefficient was
  averaged over a thermal Maxwellian distribution. In addition we also
  present elastic collision cross sections and the spin-exchange cross
  section. The rate coefficient for charge transfer was found to be
  $2.3\times 10^{-16}$ cm$^3$/sec, while those for the elastic and
  spin-exchange cross sections were found to be several orders of
  magnitude higher ($1.1\times 10^{-8}$ cm$^3$/sec and $2.3\times
  10^{-9}$ cm$^3$/sec, respectively). This confirms our assumption
  that the milli-Kelvin regime of collisional cooling of calcium ions
  by sodium atoms is favorable with the respect to low loss of calcium
  ions due to the charge transfer.
\end{abstract}

\maketitle

\section{Introduction}

A broad range of techniques from atomic physics and optics have
allowed the accurate manipulation of ultracold samples.  Recent
studies are probing ultracold atomic systems in which electric charges
may play an important role \cite{robin_na2,hopping,ion-bec}.  These
include ultracold plasmas \cite{cold-plasma1, cold-plasma2,
  cold-plasma3, cold-plasma4}, ultracold Rydberg gases
\cite{cold-rydberg1, cold-rydberg2}, as well as ionization experiments
in a BEC \cite{arimondo}.

Recently, we proposed experiments on simultaneously cooling and
trapping ions and atoms in the same confined space \cite{smith02}.
The proposed experiments will investigate the regime of cold atomic
and molecular ion-atom collision processes in a special hybrid
neutral-atom--ion trap under high vacuum collisions. The idea is to
use the techniques of laser cooling, trapping and manipulation of cold
neutral atoms to make a refrigerator for the sympathetic cooling of an
overlapping sample of atomic or molecular ions to very low
temperatures. We intend to probe fundamental collision processes and
cooling kinetics in as low a temperature range as possible.

The hybrid trap we have built consists of a conventional
magneto-optic neutral atom trap (MOT) combined with a linear Paul
radiofrequency quadruple trap. The linear Paul trap, with longitudinal
electrostatic confinement, is chosen so as to suppress most of the
micromotion and r.f. heating associated with conventional Paul traps
\cite{paultrap}.

\section{Potential curves and dipole moment}

To calculate the cross sections and rate coefficients for the charge
transfer and the elastic and spin-exchange collisions we have
performed \textit{ab initio} calculations of the adiabatic potential
curves for (NaCa)$^+$ molecular system in the X$^1\Sigma^+$,
a$^3\Sigma^+$ and A$^1\Sigma^+$ states, which go asymptotically to the
ground (Na$^+ + $Ca) and excited (Ca$^+ + $Na(3s)) states of the
ion-atom quasimolecular system. The method used was second order
M\"oller-Plesset perturbation theory (MP2) using a Gaussian triple
zeta+diffuse+polarization basis set, 6-311+G(3df). This basis set
included three sets of $d$-polarization functions and one set of
$f$-polarization functions in addition to a diffuse function on each
atomic center. The first excited A$^1\Sigma^+$ state was optimized for
the second root of a CASSCF (Complete Active Space Self-Consistent
Field) using eight orbitals in the active space. This was followed
with an MP2 (M\"oller-Plesset second order perturbation theory)
correction to the optimized CASSCF energy.

\begin{figure}
\includegraphics[width=\columnwidth]{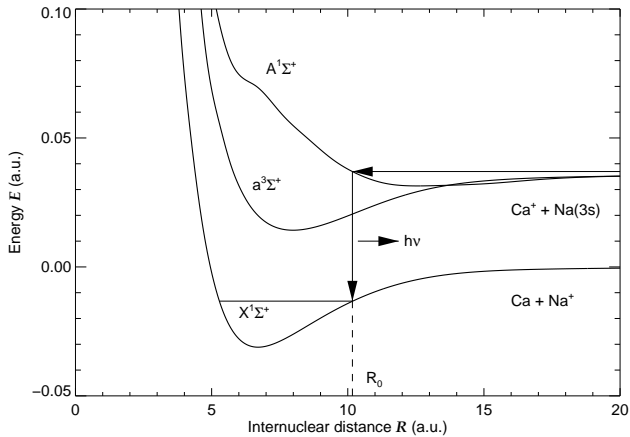}
\caption{\label{fig:pot-curves} Adiabatic potential curves for
  selected states of the (NaCa)$^+$ molecular system. $R_0$ is a
  classical turning point. Shown is a free-bound radiative transition
  with the emission of a photon $h\nu$.}
\end{figure}

Figure \ref{fig:pot-curves} shows the potential curves of the ground
and excited states of (NaCa)$^+$ molecular system in atomic units
(a.u.) 
(see also Table~\ref{tab1}).  
The curves were obtained by performing a cubic spline fit to
the calculated points. At small internuclear separations $R$, the
potential curves were approximated with the short-range potential of
the form $V(R)=(A/R) \exp(-BR)$, where $A$ and $B$ are fitting
parameters, and for $R > 30\;a_0$, the potential curves were fitted
with the appropriate long-range potentials. For the excited states,
the long-range potential was of the form
\begin{equation}
V(R) \sim  - \frac{1}{2}\left (\frac{162.7}{R^4}
             + \frac{1902}{R^6} + \frac{55518}{R^8}\right )
\label{eq:Na2}
\end{equation}
where the Na dipole (162.7 $e^2 a_0^3$),
quadrupole (1902 $e^2 a_0^5$), and
octupole (55,518 $e^2 a_0^7$) polarizabilities were taken from 
Ref.\cite{robin_na2}. 
For the ground state
X$^1\Sigma^+$, the form of the long-range
potential was
\begin{equation}
V(R) \sim - \frac{1}{2} \frac{\alpha}{R^4} ,
\end{equation}
where $\alpha = 156$ $e^2 a_0^3$ is the 
dipole polarizability of 
the calcium atom.

\begin{figure}
\includegraphics[width=\columnwidth]{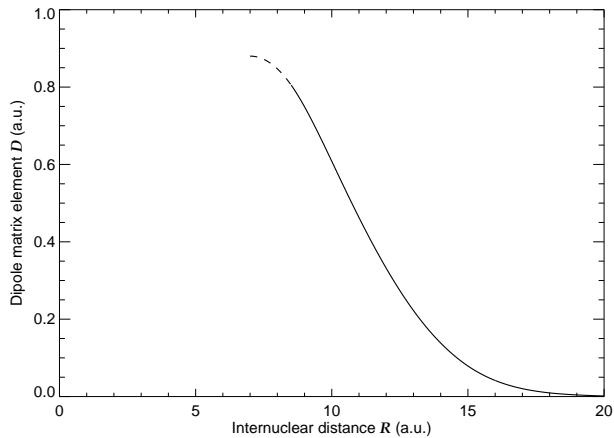}
\caption{\label{fig:dip-matrix} Calculated radial transition dipole
  matrix elements. The matrix elements are calculated at distances $R
  > 8\;a_0$ only, because the current problem did not require the
  knowledge of the matrix elements at lower inter-nuclear distances
  due to small energies of the approaching sodium atoms.}
\end{figure}

\begin{table}
\caption{\label{tab1} Potential energy and dipole matrix element values
  obtained by cubic spline fit of the adiabatic
  potential and dipole matrix curves for selected 
  states of the (NaCa)$^+$ molecular system. Potential curves are
  located relative to the ground asymptote. Notation 1.23[-4] stands for
  $1.23 \times 10^{-4}$}
\begin{ruledtabular}
\begin{tabular}{lcccr}
 & \multicolumn{3}{c}{$E$ (a.u.)}& \\
$R$ ($a_0$)& X$^1\Sigma ^+$ & A$^1\Sigma ^+$ & a$^3\Sigma ^+$ & $d$
(a.u.)\\
\hline
     3    &      3.54[-1] & 4.16[-1] & 1.25 & \\
     4    &      7.42[-2] & 1.98[-1] & 2.28[-1] & \\
     5    &     -2.15[-3] & 1.09[-1] & 6.88[-2] & \\
     6    &     -2.79[-2] & 7.47[-2] & 3.15[-2] & \\
     7    &     -3.07[-2] & 6.64[-2] & 1.73[-2] & 8.80[-1]\\
     8    &     -2.61[-2] & 5.47[-2] & 1.42[-2] & 8.47[-1]\\
     9    &     -1.99[-2] & 4.55[-2] & 1.60[-2] & 7.49[-1]\\
    10    &     -1.42[-2] & 3.79[-2] & 1.98[-2] & 6.08[-1]\\
    11    &     -9.64[-3] & 3.34[-2] & 2.39[-2] & 4.63[-1]\\
    12    &     -6.30[-3] & 3.16[-2] & 2.75[-2] & 3.32[-1]\\
    13    &     -4.01[-3] & 3.15[-2] & 3.03[-2] & 2.22[-1]\\
    14    &     -2.56[-3] & 3.18[-2] & 3.22[-2] & 1.38[-1]\\
    15    &     -1.70[-3] & 3.22[-2] & 3.33[-2] & 7.90[-2]\\
    16    &     -1.21[-3] & 3.30[-2] & 3.41[-2] & 4.15[-2]\\
    17    &     -9.17[-4] & 3.39[-2] & 3.45[-2] & 2.04[-2]\\
    18    &     -7.24[-4] & 3.45[-2] & 3.48[-2] & 9.44[-3]\\
    19    &     -5.84[-4] & 3.49[-2] & 3.51[-2] & 4.21[-3]\\
    20    &     -4.76[-4] & 3.52[-2] & 3.52[-2] & 1.53[-3]\\
\end{tabular}
\end{ruledtabular}
\end{table}

Figure \ref{fig:dip-matrix} shows the calculated radial dipole matrix
element between X$^1\Sigma^+$ and A$^1\Sigma^+$ states fitted with
cubic splines 
(see also Table~\ref{tab1}). 
The dipole matrix elements were calculated for $R >
8\;a_0$ only, since for our purposes the problem did not require the
exact knowledge of the dipole radial matrix elements at smaller
distances. At small distances there is little overlap between the
wavefunctions of the ground and excited states for the energies of the
approaching sodium atoms considered here. At milli-Kelvin energies of
the sodium atoms, the classical turning point $R_0$ was found to be
always to the right of $R=8\; a_0$.

%
% moved the section on cross sections here
%

\section{Cross sections}

A neutral atom of Na collides elastically with Ca$^{+}$ in
the singlet A$^1\Sigma^+$ or triplet a$^3\Sigma^+$ states of 
NaCa$^{+}$. The corresponding elastic cross sections are 
given by
\begin{equation}
   \sigma_{\rm el.}^{S,T} =
   \frac{4\pi}{k^{2}}\sum_{l=0}^{\infty}
   (2l+1) \sin^{2}(\eta_{l}^{S,T}) \; ,
\end{equation}
where $k=\sqrt{2\mu E}/\hbar$ with $\mu$ the reduced mass and
$E$ the collision energy in the center of mass system, 
and $\eta_{l}^{S,T}$ is the $l^{\rm th}$
partial wave phase shift corresponding to the
A$^1\Sigma^+$ and a$^3\Sigma^+$ states, respectively.
Another possible outcome of the collision between Na
and Ca$^{+}$ is spin-exchange. 
In the elastic approximation \cite{dalgarno65}, also known as
the Degenerate Internal States (DIS)
approximation \cite{verhaar}, one describes this scattering 
process in terms of the singlet and triplet scattering phase shifts,
and the spin-exchange cross section is given by
\begin{equation}
   \sigma_{\rm exch.} = \frac{\pi}{k^{2}}\sum_{l=0}^{\infty}
   (2l+1)\sin^{2}(\eta_{l}^{S}-\eta_{l}^{T}) \; .
   \label{exch}
\end{equation}
Although this approximation is valid for collision energies 
larger than the hyperfine splitting, it usually gives the right
order of magnitude at smaller energies \cite{eddy-robin}. 

The phase shifts $\eta_{l}^{S,T}$ are determined from
the continuum eigenfunctions $y^{S,T}_{E,l}(R)$, which are the
regular solutions of the partial wave equation
\begin{equation} \left( \frac{d^{2}}{dR^{2}} + k^{2}
           -\frac{2\mu}{\hbar^{2}} V_{S,T}(R)
           - \frac{l(l+1)}{R^{2}} \right) y^{S,T}_{E,l}(R) = 0 \; .
    \label{eq:free}
\end{equation}
The asymptotic form of $y^{S,T}_{E,l}(R)$ at large
distances gives the elastic scattering phase shifts
\begin{equation}
    y^{S,T}_{E,l}(R) \sim
    \sin \left[ kR - \frac{l\pi}{2} + \eta_{l}^{S,T}(k) \right] \; .
\end{equation}

\begin{figure}
\includegraphics[width=\columnwidth]{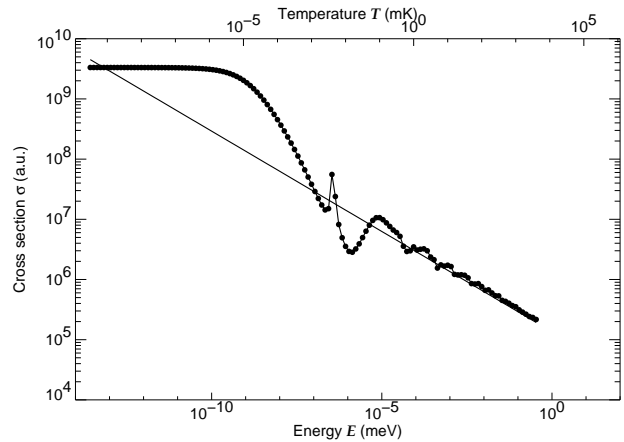}
\caption{\label{fig:cs-singlet} Elastic collision cross section for the
  A$^1\Sigma^+$ state of molecular system (NaCa)$^+$. The elastic cross
  section above 10 $\mu$K fits the approximate expression $\sigma (E)
  = 5310E^{-1/3}$, where the
  energy is in units of Hartrees and the cross section is in units of
  $a_0^2$.}
\end{figure}

\begin{figure}
\includegraphics[width=\columnwidth]{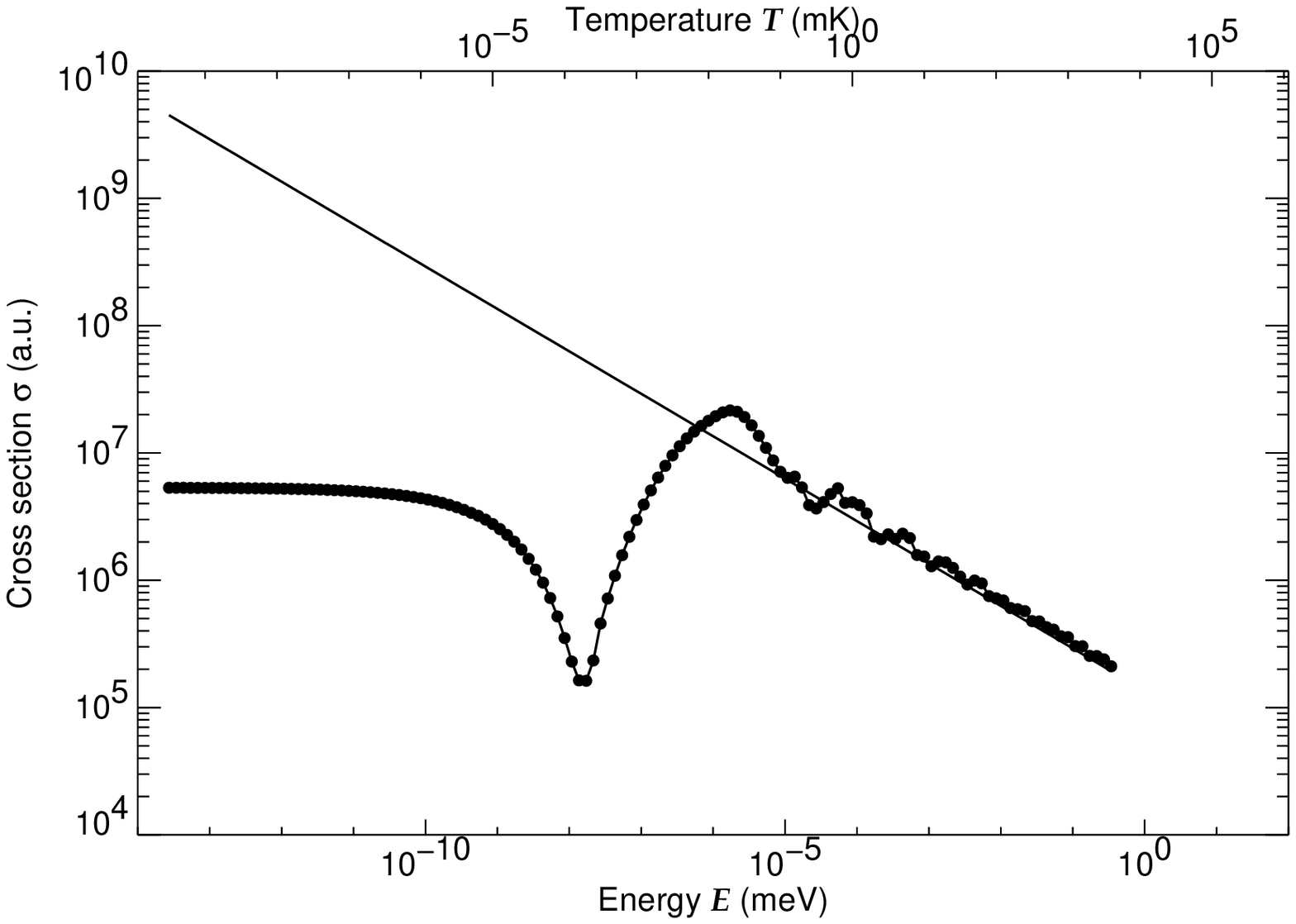}
\caption{\label{fig:cs-triplet} Elastic collision cross section for the
  a$^3\Sigma^+$ state of molecular system (NaCa)$^+$. The elastic cross
  section above 10 $\mu$K fits the approximate expression $\sigma (E)
  = 5070E^{-1/3}$, where the
  energy is in units of Hartrees and the cross section is in units of
  $a_0^2$.}
\end{figure}

\begin{figure}
\includegraphics[width=\columnwidth]{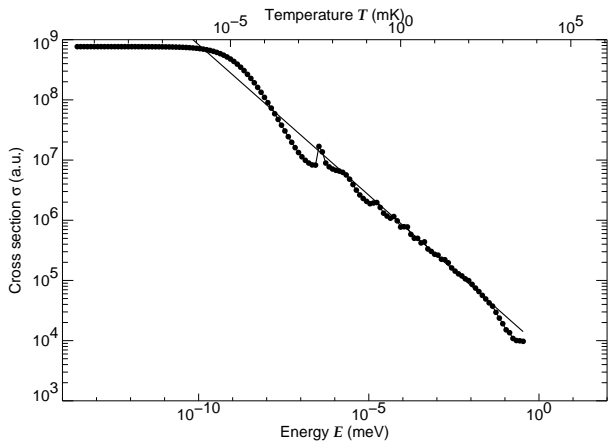}
\caption{\label{fig:cs-sf} Spin-exchange collision cross section between states
  A$^1\Sigma^+$ and a$^3\Sigma^+$ of the molecular system (NaCa)$^+$.
  The cross section fits the expression $\sigma (E) = 44E^{-1/2}$,
  where the energy is in units of Hartrees and the cross section is in
  units of $a_0^2$.}
\end{figure}

The results of our calculations are shown on 
Figs.~\ref{fig:cs-singlet} and  \ref{fig:cs-triplet}
for the singlet and triplet elastic cross sections, 
respectively, and on Fig.~\ref{fig:cs-sf} for the 
spin-exchange cross section. In all cases, the $s$-wave
contribution is dominant at energies corresponding to
temperatures below 1 nK, in sharp contrast to neutral 
alkali atoms where this take place for energies around 
100 $\mu$K \cite{review-paper:weiner}. This is due to
the very long range of the polarization potential as
compared to the shorter range of van der Waals interactions
between neutral atoms. Our {\it ab initio} 
potential curves are not of sufficient accuracy enough to predict
the $s$-wave scattering lengths associated with the singlet and triplet
potentials with a fair degree of certainty: the measurements that will
be realized in our proposed hybrid MOT would help fine-tuning the
potentials so that accurate estimates of the scattering lengths could
be given. As $E$ increases, many more partial 
waves contribute to the cross sections: already at 
$E/k_{B}\sim 1$ $\mu$K, 7 partial waves are necessary,
and over 30 are required at 1 mK. Although shape resonances
may play an important role at very small energies, their effect
is not dominant for $E/k_{B}$ larger than 100 $\mu$K
(see the smooth behavior of the cross sections in 
Figs.\ref{fig:cs-singlet} - \ref{fig:cs-sf}).
This rapid growth in the number of partial waves led us
to consider semiclassical approximations for the cross sections.
Following the method described in \cite{robin_na2}, we
found that the cross sections are well represented by
power-law functions of energy for the temperature range down 
to $\mu$K. 
Figures \ref{fig:cs-singlet} - \ref{fig:cs-sf} show 
the calculated collision cross sections together with the 
power-law fits 
\begin{eqnarray}
  \sigma_{\rm el.}(E)  & = & C_{\rm el.}E^{-1/3} \; ,\\
  \sigma_{\rm exch.}(E)& = & C_{\rm exch.}E^{-1/2} \; ,
\end{eqnarray}
where
$C_{\rm el.} = 5310$ and 5070 for the singlet
A$^1\Sigma^+$ and triplet a$^3\Sigma^+$ states, respectively, and 
$C_{\rm exch.}= 44$. These coefficients were 
obtained by performing a linear fit of the cross sections plotted 
on a log-log
scale in the energy range above 1 $\mu$K. All quantities are expressed
in atomic units. 

Using these approximate cross sections, we can calculate
the various rate coefficients $\mathcal{R}=\langle \sigma v\rangle$,
where $v=\sqrt{2E/\mu}$ is the relative velocity, and 
$\langle\cdots\rangle$ implies averaging over the 
velocity distribution. Assuming a Maxwellian velocity 
distribution characterized by the temperature $T$, we
obtain the following rate coefficients
\begin{eqnarray}
\mathcal{R}_{\rm el.}
   & = & \sqrt{\frac{8}{\pi\mu}}
\Gamma\left (\frac{5}{3}\right )(k_{B}T)^{1/6} C_{\rm el.} , \\
\mathcal{R}_{\rm exch.} 
   & = & \sqrt{\frac{2}{\mu}} C_{\rm exch.},
\label{eq:cs-el}
\end{eqnarray}
in atomic units. Note that while $\mathcal{R}_{\rm exch.}$ 
is independent of temperature, $\mathcal{R}_{\rm el.}$
varies very slowly with $T$.  
Figs. \ref{fig:cs-singlet} - \ref{fig:cs-sf}
show that the power law fits are valid at temperatures 1--10 $\mu$K 
and higher, which validates the semiclassical approach to 
calculating the rate coefficients for elastic and spin-exchange 
collisions.

As an example, suitable for the proposed experiment, we evaluated 
the rates for a temperature of 1 mK. They are
$\mathcal{R}_{\rm el.} = 1.7-1.8$ a.u. for elastic collisions
and $\mathcal{R}_{\rm exch.} = 0.38$ a.u. for spin-exchange
collisions. Converting to units cm$^3$ sec$^{-1}$, we obtain
$\mathcal{R}_{\rm el.} = (1.05-1.10)\times 10^{-8}$ cm$^3$
sec$^{-1}$ and $\mathcal{R}_{\rm exch.} = 2.33\times 10^{-9}$
cm$^3$ sec$^{-1}$.
For a typical MOT particle density $n = 10^{10}$ cm$^{-3}$ these rates
give the following values for effective elastic lifetimes $1/\tau$:
$1/\tau_{\rm el.} = \mathcal{R}_{\rm el.}n = (1.05-1.10) 
\times 10^2$
sec$^{-1}$, and 
$1/\tau_{\rm exch.} = \mathcal{R}_{\rm exch.}n = 23.3$
sec$^{-1}$. 

%
% Section on radiative charge transfer
%

\section{Radiative charge transfer}

The rate coefficient for radiative charge transfer may be 
expressed as an integral of the transition probability over the 
collision path,
averaged over the initial velocities and the impact parameters
\cite{bates51}.  
Because many partial waves contribute even
at 1 $\mu$K, we use a semiclassical treatment where the
cross section for radiative charge transfer
may be expressed as
\cite{dalg88}
\begin{eqnarray}
\sigma(E) &=&  2\pi \left ( \frac{2\mu}{E}\right)^{1/2}\nonumber\\
& & \times\int_0^\infty\!\!\!\! b \mathrm{d}b\!
\int_{R_0}^\infty\!\!\!\!\mathrm{d}R
\frac{A(R)}{[1-V(R)/E-b^2/R^2]^{1/2}},
\label{eq_sigma}
\end{eqnarray}
where $b$ is the impact parameter, $R_0$ is the classical 
turning point (distance of
closest approach), $A(R)$ is the Einstein spontaneous emission
transition probability, and $V(R)$ is the entrance channel potential
curve. 
Again, we can obtain a rate coefficient 
$\mathcal{R}_{\rm tr}=\langle \sigma v\rangle$ by averaging over a 
Maxwellian velocity distribution characterized by $T$
\cite{butler1977}
\begin{eqnarray}
  \mathcal{R}_{\rm tr}& = & 8\sqrt{\pi}\left (\frac{1}{k_{B}T}
  \right )^{3/2}
  \int_0^\infty\mathrm{d}E\int_0^\infty\mathrm{d}b\nonumber\\
           &   & \times\int_{R_0}^\infty\mathrm{d}R\frac{bE^{1/2}A(R)
                 \mathrm{e}^{-E/k_{B}T}}{[1 - b^2/R^2 - V(R)/E]^{1/2}}.
\label{eq:rate}
\end{eqnarray}
The transition probabilities are approximated by
\begin{equation}
  A(R) = \frac{4}{3}\alpha^3\omega^3(R)D(R)\; \mathrm{a.u.},
\end{equation}
where $\alpha$ is the fine structure constant, 
$D(R)$ is the dipole moment (see Fig.~\ref{fig:dip-matrix})
and
\begin{equation}
  \hbar\omega(R) = V_n(R) - V_{n'}(R)
\end{equation}
is the energy difference between the entrance and exit potential
curves (here A$^1\Sigma^+$ and X$^1\Sigma^+$, respectively),
all at a separation $R$.
Note that the effect of possible shape resonances is not taken 
into account in Eq.(\ref{eq:rate}): as mentioned in the previous 
section, shape resonances are not relevant for $E/k_{B}\sim 100$ $\mu$K or
higher (although they may play a role at much lower energies).
The integrations of the Eq. ({\ref{eq:rate}) were evaluated numerically
for temperatures ranging from 1 mK to 1000 K. The rate coefficient was
found to be independent of temperature and was practically
constant with the value of $\mathcal{R}_{\rm tr} = 2.3\times 10^{-16}$
cm$^3$/sec. With the attainable densities of ultracold sodium atoms of
the order of $n = 10^{9} - 10^{11}$ cm$^3$, the rate coefficient
gives for the the radiative lifetime of the excited A$^1\Sigma^+$
state
\begin{equation}
\tau = 1/(\mathcal{R}_{\rm tr} n) \sim 4 \times 10^4 - 4\times 10^6\;
\mathrm{sec},
\end{equation}
or in the range from several hours to several days. Parenthetically,
we note that, although charge transfer {\it without} photon emission
is possible at higher collision energies (many eV), in the 1K range
and below, the small coupling between the a and X states of (NaCa)$^+$
(see Figure \ref{fig:pot-curves}) makes the rate for nonradiative
charge transfer negligible.

We also considered the possibility of a charge transfer process that
could occur by laser excitation from the A$^1\Sigma ^+$ state of
NaCa$^+$ to a higher molecular state that correlates with Na$^+$(3s)
plus an excited Ca level. The two wavelengths that must be considered
are Na(3s $\rightarrow$ 3p) at 589.0 nm and Ca$^+$(4s $\rightarrow$
4p) at 397.0 nm. An examination of the asymptotic limits of molecular
states lying above Na(3s)$+$ Ca$^+$(4s) indicates that the closest
excited level would correspond to Na$^+$(3s) $+$ Ca[$^1$P(4p)] at
633.0 nm, too far from resonance to be of importance.\footnote{A
  second stimulated emission process could possibly arise since the
  Na$^+$ $+$ Ca[$^1$S(5s)] asymptote is close (392.7 nm) to the Ca$^+$
  resonance line at 397.0 nm. Excitation to this level, however, can
  be eliminated in the experiment by turning off the 397.0 nm ion
  cooling laser before activating the MOT.}

\section{Conclusion}

In conclusion, we have shown that the semiclassical approach to
evaluating the rate coefficients is a good approximation because many
partial waves are involved even at $\sim 1$ mK. The excited singlet
state of the colliding molecular system (NaCa)$^+$ is metastable
against radiative charge transfer with the lifetime of the order of
many hours. At the same time, the rates for elastic collisions and
also for spin exchange are much greater, which provides an efficient
mechanism for sympathetic collisional cooling of calcium ions by
sodium atoms in the meV regime and below. We are setting up an
experiment to measure collisional cooling by this mechanism.

\section*{Acknowledgments}

Work is supported in part by NSF 
grants PHY-9988215 (O.M. and W.S.) and PHY-0140290 (R.C.),
as well as the University of Connecticut Research 
Foundation (R.C.).

\bibliography{paper_1-22-03_twocol}% Produces the bibliography via BibTeX.

\end{document}